%% file: Lattice-APS-PTEP.tex
\documentclass[preprint]{ptephy_v2}

\usepackage{graphicx}
\usepackage{bm}
\usepackage{color}
\usepackage{multirow}

\begin{document}

\newcommand{\Osaka}{
  Department of Physics, The University of Osaka,
  Toyonaka 560-0043, Japan
}
\newcommand{\RIKENiThems}{
Interdisciplinary Theoretical and Mathematical Sciences Program (iTHEMS), RIKEN, Wako 351-0198, Japan
}
\newcommand{\KomabaAS}{
  Graduate School of Arts and Sciences, The University of Tokyo Komaba, Meguro-ku, Tokyo 153-8902,
Japan
}
\newcommand{\KomabaMath}{
  Graduate School of Mathematical Sciences, The University of Tokyo, Komaba, Meguro-ku, Tokyo
  153-8902, Japan
}
\newcommand{\NagoyaTagen}{
  Graduate School of Mathematics, Nagoya University, Nagoya, Japan
}
\preprintnumber{OU-HET-1267}

\title{
  $K$-theoretic computation of the Atiyah(-Patodi)-Singer index of lattice Dirac operators
}
\author{Shoto~Aoki}
\affil{\RIKENiThems}
\author{Hidenori~Fukaya}
\affil{\Osaka}
\author{Mikio~Furuta}
\affil{\KomabaMath}
\author{Shinichiroh~Matsuo}
\affil{\NagoyaTagen}
\author{Tetsuya~Onogi}
\affil{\Osaka}
\author{Satoshi~Yamaguchi}
\affil{\Osaka}

\begin{abstract}
  We show that the Wilson Dirac operator in lattice gauge theory can be identified as a mathematical object in $K$-theory and that its associated spectral flow is equal to the index. In comparison to the standard lattice Dirac operator index, our formulation does not require the Ginsparg-Wilson relation and has broader applicability to systems with boundaries and to the mod-two version of the indices in general dimensions. We numerically verify that the $K$ and $KO$ group formulas reproduce the known index theorems in continuum theory. We examine the Atiyah-Singer index on a flat two-dimensional torus and, for the first time, demonstrate that the Atiyah-Patodi-Singer index with nontrivial curved boundaries, as well as the mod-two versions, can be computed on a lattice.
\end{abstract}
\maketitle
\newpage
\section{Introduction}

The index of Dirac operators \cite{Atiyah:1968mp} has played
an important role in particle theory 
to understand nonperturbative 
nature of gauge theories.
It is closely related to chiral symmetry as well as to gauge field topology,
which is a property under continuous deformation of
fields on a continuous spacetime.
However, its formulation in lattice gauge theory, 
where chiral symmetry is broken and spacetime is discretized, 
has been a challenging problem.

It is well-known that the overlap Dirac operator \cite{Neuberger:1997fp} and 
the operator in the perfect action \cite{Hasenfratz:1998ri}
achieve an exact chiral symmetry through the Ginsparg-Wilson relation \cite{Ginsparg:1981bj}
and the associated Atiyah-Singer(AS) index is well-defined.
The formulation is, however, limited to a periodic/antiperiodic square lattice
whose continuum limit is a flat torus.

Recently the authors gave a mathematical proof
that a family of the massive Wilson Dirac operators
on an even-dimensional square lattice 
can be identified as a $K$ group element \cite{MR1043170,Karoubi}
defined on the mass parameter space
and its continuum limit given by the $\eta$ invariant
or equivalently the spectral flow  \cite{Aoki:2024sjc} \footnote{
The equivalence of the spectral flow of the
Wilson Dirac operator to the index was
empirically known in \cite{Itoh:1987iy}
and shown in \cite{Adams:1998eg} to the overlap Dirac index.
Its mathematical meaning or relation to the $K$-theory
was, however, not discussed.
}, 
converges to the index of the continuum Dirac operator.
In contrast to the standard overlap index,
our new $K$-theoretic treatment of the lattice Dirac operator
does not rely on 
the Ginsparg-Wilson relation at all, 
and offers wider applications 
to the systems where chiral symmetry
is absent or difficult to realize.

One such system is a manifold with boundaries
in which the Ginsparg-Wilson relation is generally broken \cite{Luscher:2006df}. 
The $K$-group element defined by the Wilson fermion Dirac operator can be 
straightforwardly extended to
the case with open boundary conditions,
which is equivalent to the domain-wall fermions \cite{Kaplan:1992bt,Shamir:1993zy}.
In \cite{Fukaya:2017tsq, Fukaya:2019qlf}, in which five of the authors
were involved, it was proved in continuum theory that the
Atiyah-Patodi-Singer (APS) index \cite{Atiyah:1975jf} 
of the Dirac operator is equal to
the $\eta$ invariant of the massive Dirac operator
where the mass term has domain-walls at the location of
the original boundaries.
A perturbative equivalence between the continuum
and lattice $\eta$ invariants was examined in Ref. \cite{Fukaya:2019myi}
and a more rigorous mathematical justification will be given in \cite{Fukaya:2025prep}.

Another example is the mod-two index \cite{Fukaya:2020tjk}
of real Dirac operators, which appears in arbitrary dimensions.
A well-known example is the one in the $SU(2)$ gauge theory in
five dimensions, which is known as the origin of the Witten anomaly \cite{Witten:1982fp}.
In odd dimensions, chiral symmetry is absent, and
the overlap Dirac operator is not directly available.
In contrast, $K$-theory still works with 
modifications that are well-known in mathematics:
switching from $K$ groups to $KO$ groups.

These results are nontrivial in mathematics 
(for different mathematical approaches, we refer to Refs.~\cite{Yamashita:2020nkf,Kubota:2020tpr}).
The index of the Dirac operator is generally formulated and nontrivial only when
the system has an infinite-dimensional Hilbert space.
Our formulation may lead to a systematic way 
of a finite-dimensional approximation of topological invariants of 
general vector bundles.
We also remark that our $K$-theoretic formulas
for the APS index provide a natural connection between
the massive bulk and the massless edge-localized fermions,
which may be useful for understanding topological insulators and superconductors.

In this work, we numerically verify that our
$K$-theoretic formulas yield values  consistent 
with the index theorems in continuum theory.
On a two-dimensional square lattice, we first examine
the Atiyah-Singer index theorem for a $U(1)$ gauge theory
on a flat torus $T^2$.
Then we introduce nontrivial domain-walls, 
including curved ones \cite{Aoki:2022cwg,Aoki:2022aez,Aoki:2023lqp,Kaplan:2023pxd,Kaplan:2023pvd,Aoki:2024bwx,Clancy:2024bjb,Kaplan:2024ezz}, 
and investigate the Atiyah-Patodi-Singer index theorem,
in which the edge-localized modes at the domain-wall
play a crucial role.
We also study a Majorana fermion system with a boundary
to test the mod-two index formula. 
To the best of our knowledge, this is the first numerical study
demonstrating that the APS index and its mod-two version 
can be computed on a lattice.

\section{$K$-theoretic formulas}

Here we summarize the index formulas
obtained from $K$-theory.
For mathematical details,  see \cite{Aoki:2024sjc,Fukaya:2025prep}.
We also provide a brief summary in Appendices.~\ref{app:K} and \ref{app:KO}.

In the standard formulation of the index in continuum theory,
the massless Dirac operators are interpreted
to represent the $K^0(\{0\})$ or $KO^{-1}(\{0\})$ group elements.

The basic idea is to add and vary the mass parameter 
by $-sM$ where $M$ is positive,
and $s\in I=[-1,1]$. We denote its two end points by $\partial I=\{-1,1\}$.
The original zero eigenvalues of the massless Dirac operator
cross zero along the path $I$.
Thus the index can be evaluated 
appropriately counting the crossing-zero modes, or the spectral flow.
The equivalence is mathematically guaranteed by the so-called
suspension isomorphism between the $K^0(\{0\})$ and $K^1(I,\partial I)$ groups
(or $KO^{-1}(\{0\})$ and $KO^0(I,\partial I)$ groups for real Dirac operators.).

For a general complex Dirac operator 
we take $\gamma(D- sM)$ with $s\in I$, as
a $K^1(I,\partial I)$ group element
  under a condition that $\gamma(D \pm M)$ have no zero mode.
  Here the chirality operator $\gamma$ and its anticommutation relation with $D$ are not
  essential and we do not have to distinguish the continuum and lattice Dirac operators in the formulation.
Therefore, the massive Wilson Dirac operator (family) $\gamma(D_W- sM)$ with $0<M<2$ (to avoid fermion doubling) on an even-dimensional periodic square lattice 
  can also be identified as an element of $K^1(I,\partial I)$.
  Denoting the number of zero-crossing modes from positive to negative by $n_+$
  and that from negative to positive by $n_-$, the spectral flow
\begin{equation}
n_+-n_-= - \frac{1}{2}\eta(\gamma(D_W-M))+\frac{1}{2}\eta(\gamma(D_W+M)),
\end{equation}
is well-defined and agrees with the overlap Dirac index.
Here the eta invariant $\eta(H)$ is difference between the positive
and negative eigenvalues of a Hermitian operator $H$.
Its equivalence to that in continuum theory
at a sufficiently small lattice spacing was proved in \cite{Aoki:2024sjc}.

It is straightforward to extend the formulation
to the domain-wall fermion case.
Let us consider a flat torus $T^{2n}$
and divide it into a closed region $T_-\subset T^{2n}$ and $T_+=T^{2n}\setminus T_-$.
With the same parameter $s\in I$,
we set the mass term by
\begin{align}
  M_s(x)=\left\{
  \begin{array}{cc}
    M & \mbox{for $x\in T_+$}\\
   -sM & \mbox{for $x\in T_-$}
  \end{array}\right.,
\end{align}
and extend it on a lattice in an obvious way.
Using the Wilson Dirac operator with this domain-wall mass term,
$\gamma(D_W+ M_s)$ can be treated as a $K^1(I,\partial I)$ group element under 
the condition that there is no zero mode at $s=1$.
The spectral flow expression of the index
\begin{equation}
\label{eq:formulaAPS}
n_+-n_- = -\frac{1}{2}\eta(\gamma(D_W+M_{+1}))
+\frac{1}{2}\eta(\gamma(D_W+M_{-1})),
\end{equation}
is well-defined. 
In \cite{Fukaya:2019qlf}, it was proved in continuum theory
that under reasonable conditions such as sufficiently large mass value,
the $\eta$ invariant of the domain-wall Dirac
operator equals to the APS index on $T_-$.
Its mathematical equality to the lattice version 
will be shown in \cite{Fukaya:2025prep}.

According to the APS index theorem \cite{Atiyah:1975jf}, 
the index is related to a geometric quantity.
For a flat two-dimensional torus $T^2$, it is expressed as
\begin{equation}
 Q=\frac{1}{2\pi}\int_{T^2} F-\frac{1}{2}\eta(-iD_{\partial T_-}),
\end{equation}
where $F$ is a curvature two-form and 
the second term is the $\eta$ invariant of
the boundary Dirac operator $D_{\partial T_-}$ on the 
domain-wall $\partial T_-$  between $T_+$ and $T_-$.


Finally let us discuss an application to the real Dirac operators.
When the massless Dirac operator $D$ 
is real, for any eigenfunction $D\phi = i\lambda \phi$,
its complex conjugate $\phi^*$ is another eigenfunction with
the opposite signed eigenvalue $-i\lambda$.
Therefore, the number of the zero modes mod 2 is
a topological invariant, which is known as the mod-two index.
In order to express this index by the spectral flow,
we consider two flavor system to make an anti-Hermitian Dirac operator,
for example, given by
\begin{equation}
 A_s = \sigma_1 \otimes D + i\sigma_2 \otimes M_s,
\end{equation}
where $M_s$ denotes the one parameter family $s\in I$ 
of the mass with or without domain-wall.
Since this two-flavor Dirac operator anticommutes with 
$\sigma_3\otimes 1$, $A_s$ is an element of  
the $KO^0(I,\partial I)$ group.

There exists another type of the mod-two index
where an additional symmetry makes the zero modes to
appear in pairs.
The number of such pairs mod 2 is a topological invariant,
and the corresponding spectral flow characterizes
the $KO^{-1}(I,\partial I)$ group.
In this case, we do not need to increase the flavors
and the massive operator family $A_s$ 
is obtained in a more straightforward way. 
These two types of the mod-two spectral flows
are treated in a uniform way.
This is done by a forgetful map
from $KO^0(I,\partial I)$ to $KO^{-1}(I,\partial I)$
neglecting the anti-commutation $\{A_s,\sigma_3\otimes 1\}=0$.


Since the eigenvalues of $A_s$ are $\pm$ symmetric,
any crossing zero occurs in such pairs and therefore, we do not distinguish
$n_+$ and $n_-$ but count $n$ mod 2 of such pair crossings.
As shown in \cite{Fukaya:2019qlf} the mod-two version of the spectral flow is no more expressed by
the $\eta$ invariant but given by the determinant of the Dirac operators.
For the lattice Wilson Dirac operator it is given by 
\begin{equation}
\label{eq:formulaAPSmod2}
 n = \frac{1-\mbox{sgn}\det[(D_W+M_{+1})/(D_W+M_{-1})]}{2}\;\;\;\mbox{mod 2}.
\end{equation}

We emphasize here that the above lattice formulas 
(\ref{eq:formulaAPS}) and (\ref{eq:formulaAPSmod2})
were not known in the literature and become nonzero 
in the systems where the Ginsparg-Wilson relation 
is absent or not available\footnote{
It will be interesting to compare our formulation with a
recent study \cite{Clancy:2023ino,Singh:2025wet} where
the Ginsparg-Wilson relation was generalized to
discrete symmetries so that the mod-two type index
can be defined.
}.


\section{Numerical lattice results}

In this section, we numerically verify on a two-dimensional square lattice
that the $K$-theoretic formulas in the previous section
using the lattice Wilson Dirac operator,
are consistent with the continuum Atiyah-Singer and
Atiyah-Patodi-Singer index theorems.
Here we set the lattice spacing $a=1$.

The two-dimensional Wilson Dirac operator we consider is
\begin{equation}
 D_W+M_s = \sum_{i=1,2} \left[\sigma^i\frac{\nabla^f_i+\nabla^b_i}{2} \textcolor{black}{-\frac{1}{2}\nabla^f_i \nabla^b_i} \right]+M_s,
\end{equation}
with 
\begin{equation}
 \sigma_1=\left(\begin{array}{cc}
	   0 & 1\\ 1 & 0
		\end{array}\right),\;\;\;
\sigma_2=\left(\begin{array}{cc}
	   0 & -i\\ i & 0
		\end{array}\right),
\end{equation}
and the covariant forward and backward difference at a position $\bm{x}=(x,y)$
are given by
\begin{align}
\nabla^f_i\psi(\bm{x})&=U_i(\bm{x})\psi(\bm{x}+\bm{e}_i)-\psi(\bm{x}),\nonumber\\
\nabla^b_i\psi(\bm{x})&=\psi(\bm{x})-U^\dagger_i(\bm{x}-\bm{e}_i)\psi(\bm{x}-\bm{e}_i),
\end{align}
where $\bm{e}_i$ is the unit vector in the $i$-th direction
and $U_i(\bm{x})$ is the corresponding gauge link variable.
This operator is complex in general but when we multiply the
chirality operator
\begin{equation}
\gamma=\sigma_3=\left(\begin{array}{cc}
	   1 & 0\\ 0 & -1
		\end{array}\right),
\end{equation}
we obtain a Hermitian operator $\gamma(D_W+M_s)$.

\subsection{Atiyah-Singer index on a flat torus}

First, we set the lattice size $L=33$ and impose periodic boundary
condition in the $x$ direction while  anti-periodic boundary condition
in the $y$ direction is put in order to avoid accidental zero modes.

\begin{figure}[tbhp]
  \centering
  \includegraphics[width=0.5\columnwidth]{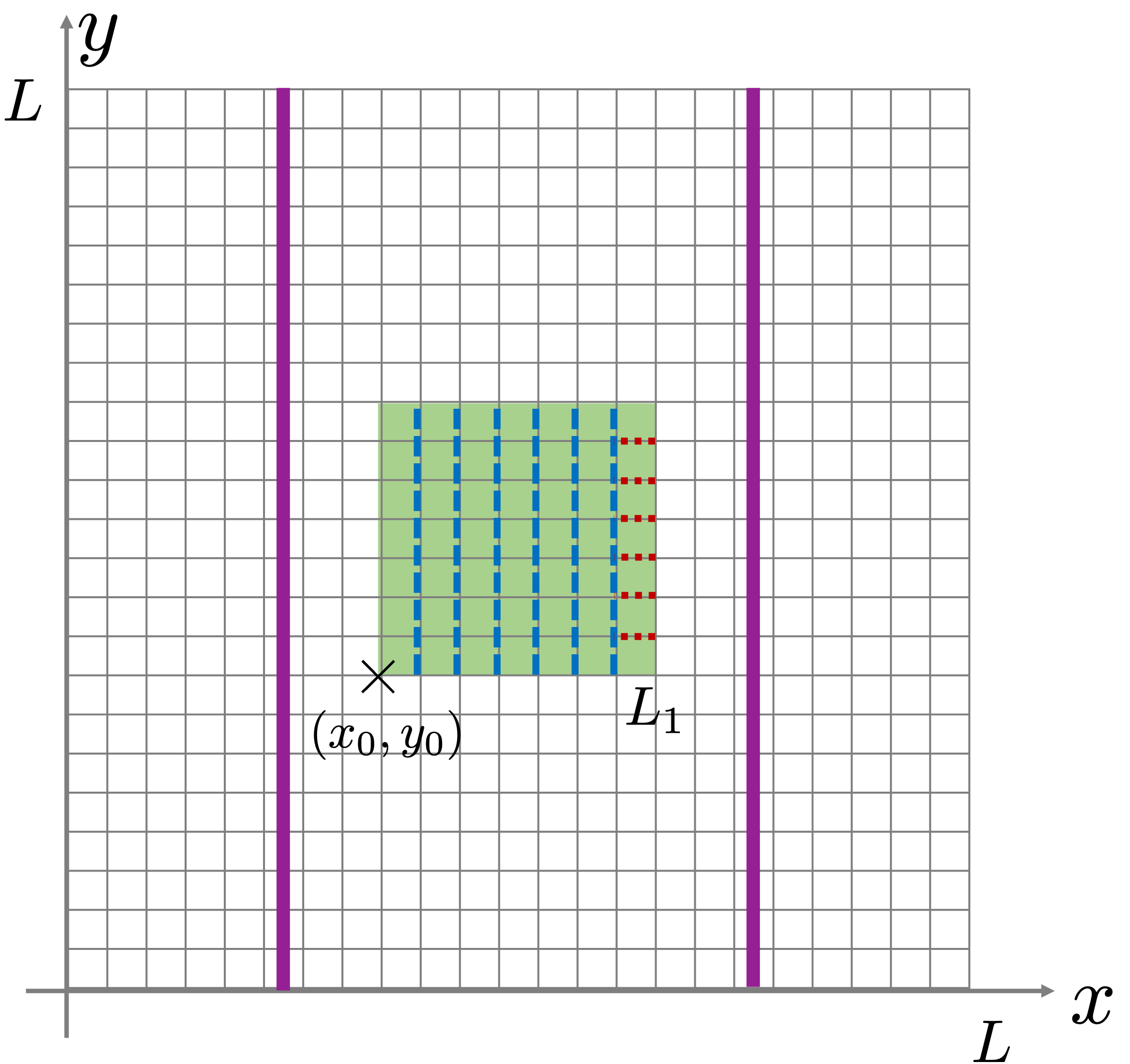}
  \caption{
    The lattice setup for the Wilson Dirac operator with and without 
flat two domain-walls.
The two thick vertical lines show the location of the 
two domain-walls when they are needed.
    We impose the periodic boundary condition in the $x$ direction
while the $y$ direction is anti-periodic.
We assign non-trivial $U(1)$ link variables to those 
depicted by the dashed and dotted lines
so that the constant plaquettes inside the shaded $L_1\times L_1$
region give the topological index $Q$.
}
    \label{fig:setup}
  \centering
  \includegraphics[width=0.7\columnwidth]{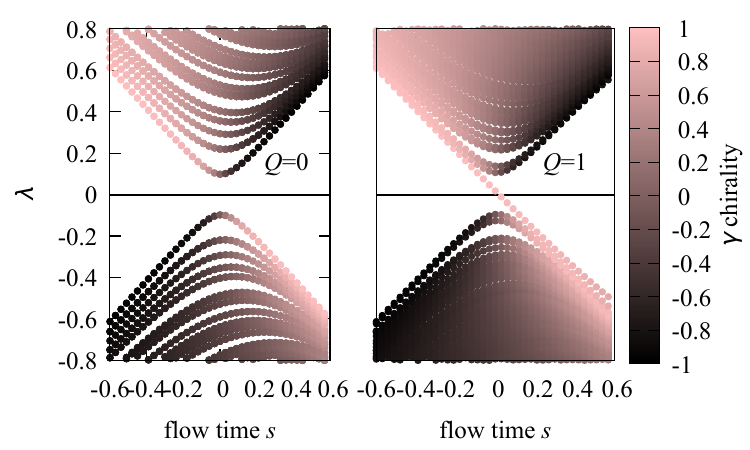}
  \caption{
    The eigenvalue spectrum of the massive Wilson Dirac operator $\gamma(D_W-Ms)$ with $M=1$.
We put a periodic boundary condition
    in the $x$ direction, while the $y$ direction is anti-periodic.
    The gradation indicates the expectation value of the
     chirality operator $\sigma_3$.
      For the left panel we give the geometrical index $Q=0$,
      while for the right panel $Q=1$.
     The spectral flow agrees with the Atiyah-Singer index theorem on a torus.
  }
    \label{fig:AS}
\end{figure}

On this lattice, we consider a sublattice of $L_1\times L_1$ square 
with $L_1<L$, whose left-bottom corner is located at $(x_0,y_0)$.
We assign nontrivial $U(1)$ gauge link variables
inside the $L_1\times L_1$ square region shown in Fig.~\ref{fig:setup},
or the link variables indicated by dotted and dashed lines
(note that all the links at the boundary of the square are taken trivial)
\begin{align}
  U_y(x,y) &=e^{\frac{2\pi i Q(x-x_0)}{L_1^2}} \; \mbox{for $0<x-x_0<L_1$ and $0<y-y_0<L_1$},\nonumber\\
  U_x(x_1,y) &=e^{\frac{-2\pi i Q(y-y_0)}{L_1}} \; \mbox{for $0<y-y_0<L_1$},
\end{align}
where $x_1=x_0+L_1-1$ and $U_{x,y}(x,y)=1$ otherwise.
Then, every plaquette inside the square is $\exp(i2\pi Q/L_1^2)$
and the geometrical index we have in the classical continuum limit \cite{Fukaya:2003ph} is 
\begin{equation}
 \frac{1}{2\pi}\int_0^L  dx \int_0^L dy F_{12} = L_1^2\times \frac{Q}{L_1^2}=Q.
\end{equation}

We numerically solve the eigenvalue problem of $\gamma(D_W-Ms)$ varying the flow time $s\in [-1,1]$,
where $\gamma$ 
is the chirality operator.
We set  $M=1$, $L_1=9$ and $(x_0,y_0)=(12,12)$.
In Fig.~\ref{fig:AS} we plot the eigenvalue spectrum as a function of $s$.
The gradation indicates the expectation value of the
chirality operator $\gamma$.
For the left panel we give the geometrical index $Q=0$,
while for the right panel $Q=1$.
The spectral flow agrees well with the Atiyah-Singer index theorem.

\subsection{Atiyah-Patodi-Singer index on a cylinder}

\begin{figure}[tbhp]
   \includegraphics[width=0.7\columnwidth]{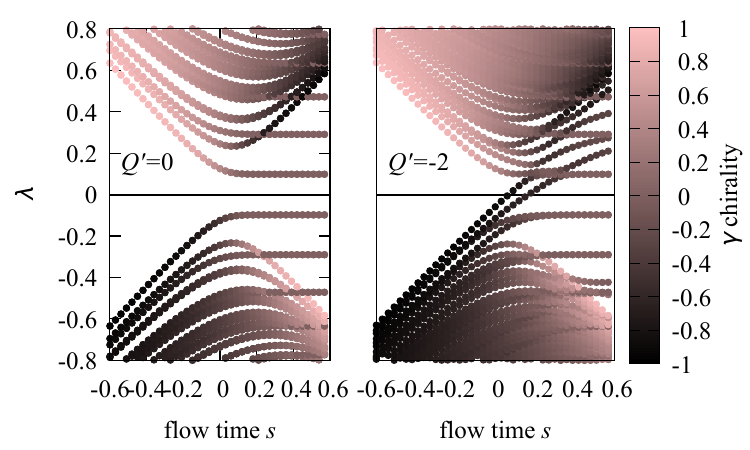}
\centering
   \caption{
     The eigenvalue spectrum of the domain-wall Dirac operator
     where two flat domain-walls are put for $s>0$.
     For the left panel we give the $U(1)$ gauge flux $Q'=0$,
     while for the right panel $Q'=-2$.
     For $s>0$, the edge-localized modes appear between $\pm sM$.
     The spectral flow agrees with the APS index theorem
     on a cylinder between the two domain-walls.
  }
  \label{fig:APSflat}
\end{figure}

Second, we vary the mass term $-sM$ only in the region $10\le x \le 22$
while it is kept at $M$ outside.
The same boundary conditions are imposed as the first case.
For $s>0$, we have two domain-walls at $x=10,22$ as illustrated in Fig.~\ref{fig:setup}
and the negative mass region forms a cylinder with
the two $S^1$ boundaries.
The eigenvalue spectrum of this domain-wall Dirac operator
is plotted in Fig.~\ref{fig:APSflat}.     
For the left panel we give the $U(1)$ gauge flux $Q'=\frac{1}{2\pi}\int F=0$,
while for the right panel $Q'=-2$.
For $s>0$, the edge-localized modes appear between the mass gap $\pm sM$.
Since the edge-localized spectrum is almost $\pm$ symmetric,
the $\eta$ invariant of the one-dimensional Dirac operator is expected to be zero.
The spectral flow agrees with the APS index theorem
on a cylinder between the two domain-walls.

\subsection{Atiyah-Patodi-Singer index on a disk}

\begin{figure}[tbhp]
  \centering
  \includegraphics[width=0.5\columnwidth]{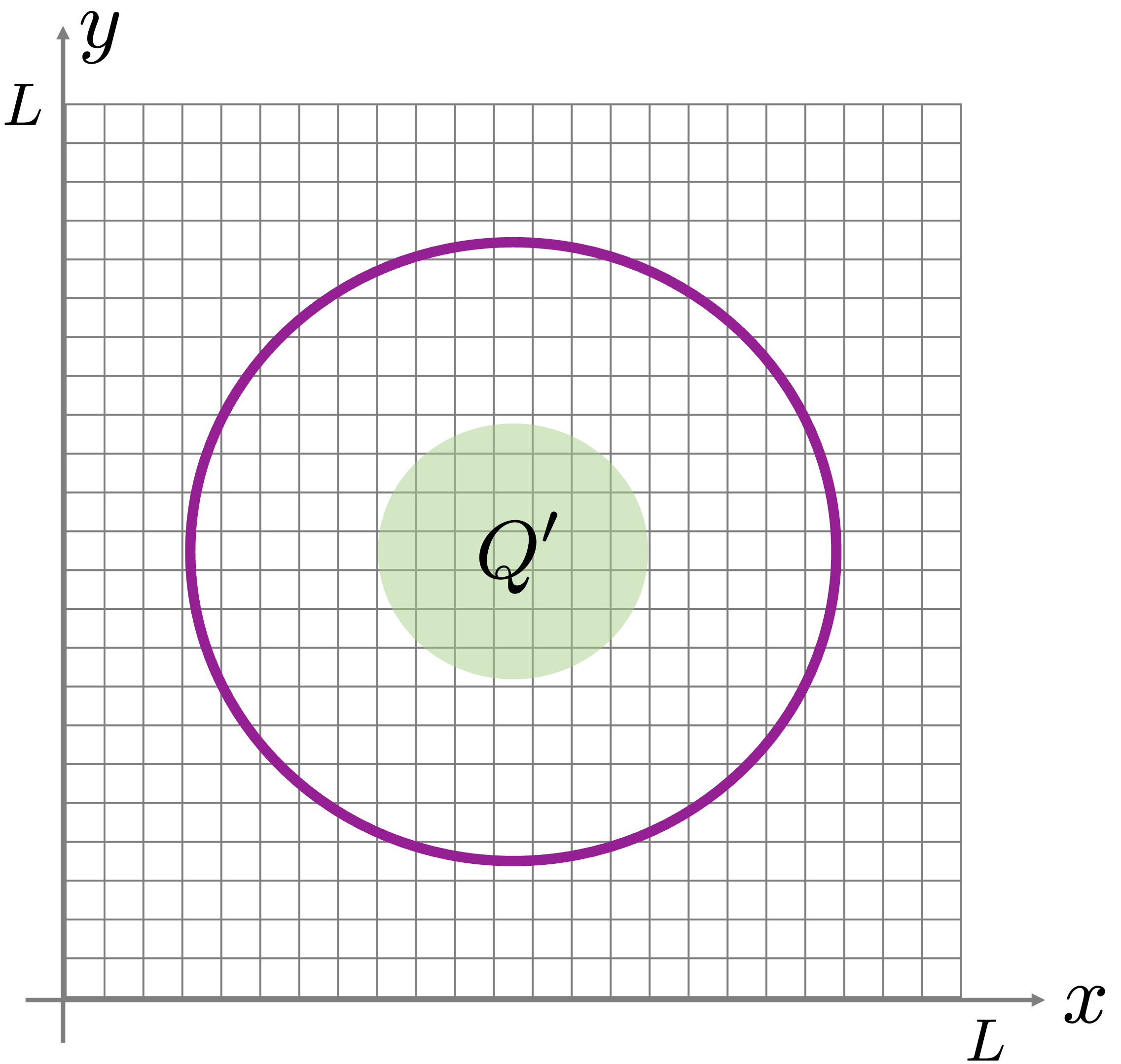}
  \caption{
    The lattice setup for the Wilson Dirac operator with a circular domain-wall.
    We impose the periodic boundary condition in the $x$ direction
while the $y$ direction is anti-periodic.
We assign non-trivial $U(1)$ link variables within the shaded circle
so that the constant plaquettes inside give the total flux $Q'$.
}
    \label{fig:setup2}
  \includegraphics[width=0.7\columnwidth]{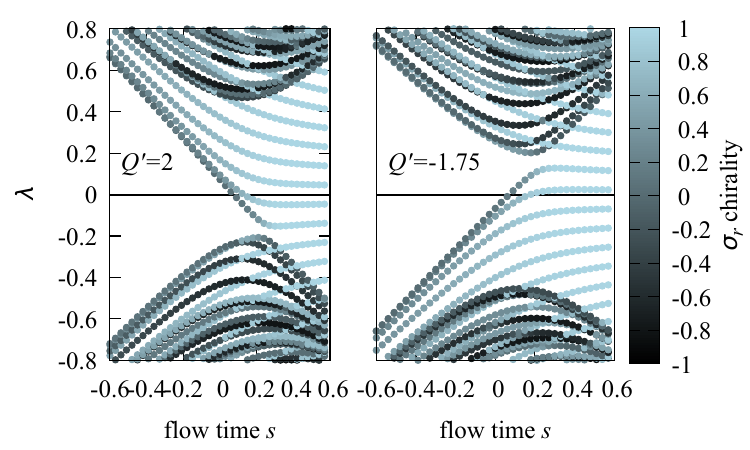}
  \caption{
    The eigenvalue spectrum 
    where a circular domain-wall is put for $s>0$.
    The gradation denotes the expectation value of $\sigma_r$,
    which corresponds to the chirality operator of the edge-localized
    modes on the circle.
    For the left panel we give the $U(1)$ gauge flux $Q'=2$,
    while for the right panel we assign a non-integer value $Q'=-1.75$,
    which are consistent with the estimated APS indices.
  }
  \label{fig:APScircle}
  \end{figure}

Third, as a more nontrivial setup with a curved boundary, 
we consider a circular domain-wall with radius $r_0=10$ as shown in Fig.~\ref{fig:setup2}.
In the region $\sqrt{x^2+y^2}<r_0$ we vary the mass $-sM$ 
in the range $s\in [-1,1]$, while it is kept at $+M$ outside.
On the internal disk, we consider the $U(1)$ gauge potential given by
\begin{align}
 A_x(x,y)&=\left\{
\begin{array}{cc}
 -\frac{Q'y}{r_1^2}& (\sqrt{x^2+y^2}<r_1) \\
 -\frac{Q'y}{x^2+y^2} & (\sqrt{x^2+y^2}\ge r_1)
\end{array}
\right.,\nonumber\\
 A_y(x,y)&=\left\{
\begin{array}{cc}
 \frac{Q'x}{r_1^2}& (\sqrt{x^2+y^2}<r_1) \\
 \frac{Q'x}{x^2+y^2}& (\sqrt{x^2+y^2}\ge r_1)
\end{array}
\right.,
\end{align}
where $r_1$ is a fixed radius in which 
the constant curvature $2Q'/r_1^2$ is given
and $Q'$ equals to the total flux inside the domain-wall.
Then we define the link variables by 
\begin{align}
U_x(x,y)&=\exp \left[i\int_{x}^{x+a} dx' A_x(x',y)\right],\nonumber\\
U_y(x,y)&=\exp \left[i\int_{y}^{y+a} dy' A_y(x,y')\right].
\end{align}
Because of the periodic or anti-periodic boundary conditions
in the $x$ and $y$ directions, the curvature is
non-trivial around $x\sim L$ and $y\sim L$ 
but that region should not affect the index 
expressed by the spectral flow when 
the mass term only inside the domain-wall is changed.

The spectrum of the circular domain-wall Dirac operator
is plotted in Fig.~\ref{fig:APScircle}.
In this case, we measure the expectation value of
the gamma matrix in the radial direction we denote by 
\begin{equation}
 \sigma_r = \frac{(x-x_0)\sigma_1+(y-y_0)\sigma_2}{\sqrt{(x-x_0)^2+(y-y_0)^2}}.
\end{equation}
which is represented by gradation of the symbols.
Note that for $s>0$ the edge-localized modes appear
between $\pm sM$ and they are the eigenstates of $\sigma_r=1$.
For the left panel we give the $U(1)$ gauge flux $Q'=2$ inside $r<r_1$,
while for the right panel we assign a non-integer value $Q'=-1.75$.
For the latter, the spectral flow is $-$2.

The non-integer part of $Q'$ must be cancelled 
by that of the $\eta$ invariant of the boundary Dirac operator,
which can be estimated from the asymmetry of the 
edge-localized mode spectrum.
In \cite{Aoki:2022aez}, the Dirac eigenvalue in the same domain-wall setup in continuum theory 
was computed as
\begin{equation}
 \lambda_j = \frac{1}{r_0}\left(j+\frac{1}{2}-Q'\right),
\end{equation}
where the integer $j$ represents the angular momentum, which makes
the interval of the eigenvalues a constant $1/r_0$.
Note that competition between the gravitational effect $1/2$ due to the curve of the domain-wall 
and the Aharonov-Bohm affect given by $Q'$ determines the asymmetry.
When $Q'$ is an integer, the spectrum is $\pm$ symmetric, which is consistent with
our numerical values of the two nearest-zero eigenvalues  
$-$0.0462492 and 0.0462905 at $Q'=2$ and $s=0.6$.
When $Q'=-1.75$, the two nearest-zero eigenvalues in continuum theory should be
$\lambda_{-2}=0.025$ and $\lambda_{-3}=-0.075$, respectively, and the asymmetry is 
$(\lambda_{-2}+\lambda_{-3})/2\lambda_{-2}=-1.0$.
Our corresponding numerical data 0.023118, $-$0.0694144 and $-$1.00131
are consistent with these\footnote{
The ratio $(\lambda_{-2}+\lambda_{-3})/2\lambda_{-2}$ agrees with the continuum theory 
better than the eigenvalues themselves. This may imply some effective shift of the 
domain-wall radius $r_0$ on the lattice.
}. In both cases, the interval of the eigenvalues look constant at $s=0.6$.
In \cite{Fukaya:2017tsq}, the $\eta$ invariant on the circle was computed
as $-\eta(-iD_{1D})/2=[Q']-Q'$ where $[\cdots]$ denotes the Gauss symbol, or the largest integer $\le Q'$.
For $Q=2$, it is zero, while for $Q'=-1.75$, it is $-$0.25.
The corresponding values $2$ and $-2$ of the spectral flow in the above two cases 
agree well with the continuum estimates for the APS index on the disk.

\subsection{Mod-two Atiyah-Patodi-Singer index on a disk and $T^2$ with an $S^1$ hole}

Finally, we change the boundary condition to periodic
in both of the $x$ and $y$ directions and consider 
a free massive Wilson Dirac operator
setting all the link variables unity.
In the same way as the previous example,
we put the circular domain-wall with radius $r_0=10$ for $s>0$.
In this case, the Wilson Dirac operator has a real structure since there is
an anti-unitary operator $C=\sigma_1 K$
where $K$ takes the complex conjugate and $C^2=1$,
\begin{equation}
 [i\gamma(D_W+M_s)] = C [i\gamma(D_W+M_s)] C^{-1}.
\end{equation}
Therefore, we can interpret it as the Dirac operator
of Majorana fermions.

\begin{figure}[tbhp]
\includegraphics[width=0.7\columnwidth]{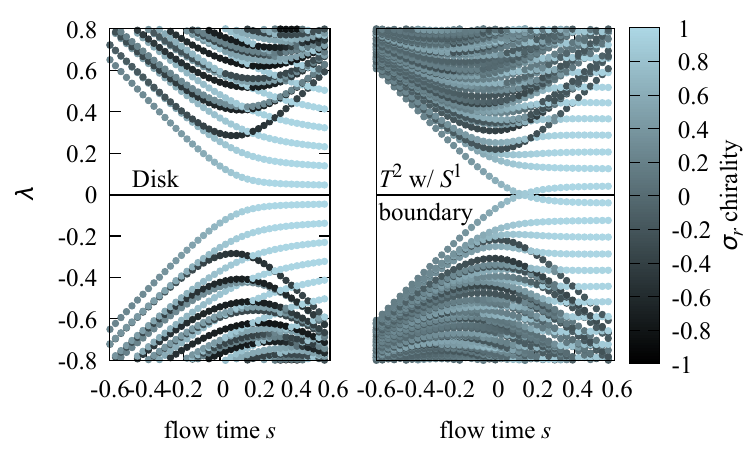}
\centering
  \caption{
    The eigenvalue spectrum of the free domain-wall Dirac operator
    where a circular domain-wall with radius $r_0=10$ is put for $s>0$.
    We assign the periodic boundary condition in both of the
    $x$ and $y$ directions.
    For the left panel, we vary the mass inside the wall $r<r_0$ by $-sM$,
    while it is kept at $+M$ outside. 
    For the right panel, we take the opposite way:
    the mass is fixed at $+M$ inside $r_0$ and
    set $-sM$ inside. The mod-two spectral flow agrees with
the mod-two APS index.
  }
  \label{fig:mod2APS}
\end{figure}

When the domain-wall is absent, the index is trivially $n=1$
since the $p_x=p_y=0$ point has two-fold degeneracy due to the spin degrees of freedom.
In the massive case, the corresponding pair of zero crossing modes
appear in the spectrum between $-M<m<M$.
In contrast, the domain-wall makes the system nontrivial.
The left panel of Fig.~\ref{fig:mod2APS} shows the domain-wall Dirac operator spectrum,
for which we vary the mass inside the wall $r<r_0$ by $-sM$, while it is kept at $+M$ outside. 
For the right panel, we take the opposite way: the mass is fixed at $+M$ inside $r_0$ and
set $-sM$ outside.
The mod-two spectral flow, which counts the number of pairs which cross zero,
agrees with the
continuum prediction of the mod-two APS index $n=0$
on the disk  \cite{Falomir:1996as} for the left panel.
The right panel indicates $n=1$ agreeing with the mod-two APS index on the $T^2$ with an $S^1$ hole,
which can be understood from the gluing property: the two cases sum up to $n=1$ of the whole $T^2$.

\section{Summary}

We have identified the massive Wilson Dirac operator in lattice gauge theory
with or without domain-walls as a mathematical object in $K$-theory.
The $K^1(I,\partial I)$ and $KO^0(I,\partial I)$ groups
and the associated spectral flows 
offer a unified formulation of various types of Dirac indices on
a discretized lattice.
Our new formulation
does not require chiral symmetry or
the Ginsparg-Wilson relation, unlike the
standard formulation of the overlap Dirac operator index.
It has a natural extension to the
fermion systems on a manifold with boundaries which may be curved,
and to those with a real structure in arbitrary dimensions.

The mathematical justification in Refs.~\cite{Aoki:2024sjc,Fukaya:2025prep}
shows that, for a sufficiently small lattice spacing,
the $K$-theoretic formulas of the Wilson Dirac operator
agree with the continuum indices under reasonable conditions.
In this work, we have examined whether it is achieved at
a numerically accessible level, using a two-dimensional square lattice of size $L=33$.

First, we have examined the Atiyah-Singer index theorem on a flat torus $T^2$.
By taking a $U(1)$ gauge field to form a constant background
in a limited region, we have confirmed 
that the spectral flow of the Wilson Dirac operator agrees well with
the estimated geometrical index $Q$.
Next, we have introduced nontrivial domain-walls, both flat and curved,
and investigated the Atiyah-Patodi-Singer index theorem.
In this case, the edge modes have been shown to localize at the domain-wall
and yield a nontrivial $\eta$ invariant, which compensates for
the non-integer part of the gauge flux $Q'$.
Finally, we have investigated the Majorana fermion system
and found that the mod-two index on the disk and
$T^2$ with an $S^1$ hole is well described
by the number of eigenvalue pairs crossing zero.

In our numerical setup, the link variables give a rather rough
approximation of the corresponding continuum gauge configuration
compared to what employed for the mathematical proof
\cite{Aoki:2024sjc, Fukaya:2025prep}, 
where the relation between the continuum gauge field and 
the lattice approximation are more carefully chosen.
We also note that we have used no assumption on the domain-walls
given on our lattice, whereas in the continuum setup  
for the boundary or the domain-wall in \cite{Fukaya:2019qlf}, 
we assumed that the metric and gauge field on 
a collar neighborhood of the boundary are flat.
In spite of the simple and crude implementation, 
our numerical results with $L=33$, $L_1=9$, $r_0=10$, $r_1=6$ in this work
show a good agreement between the index expressed by 
the spectral flow and the estimated value in continuum theory.
\\
\par
We thank Yoshio Kikukawa, Yosuke Kubota for useful discussions.
This work was supported by JSPS KAKENHI Grant Numbers 21K03222, JP21K03574, JP22H01219, JP23K03387, JP23K22490 and JP23KJ1459.


\input{suppl-PTEP.tex}

\bibliographystyle{ptephy}
\bibliography{ref.bib}

\end{document}

%% file: suppl-PTEP.tex
\begin{appendix}

\section{$K^0(\{0\})$ and $K^1(I,\partial{I})$ groups}
\label{app:K}

The standard definition of the Dirac operator index 
requires a massless Dirac operator with chiral symmetry. 
It naturally appears in the $K^0(\{0\})$ group.
Here $\{0\}$ denotes a point.

The elements of the $K^0(\{0\})$ group are given by
the Hermitian Dirac operator $\gamma D$ and its Hilbert space $\mathcal{H}$,
and the chirality operator $\gamma$.
We denote them by $[\mathcal{H},\gamma D,\gamma]$, where
$[\cdots]$ is understood taking an equivalence class.
Two triples $(\mathcal{H}_1,\gamma_1 D_1,\gamma_1)$
and $(\mathcal{H}_2,\gamma_2 D_2,\gamma_2)$ are equivalent when 
there exists another triple $(\mathcal{H}_3,\gamma_3 D_3,\gamma_3)$
with an invertible operator $\gamma_3 D_3$ and a one-parameter family of a 
larger Dirac operator $H_t (t\in [0,1])$ acting on
$\mathcal{H}_1 \oplus \mathcal{H}_2 \oplus \mathcal{H}_3$
such that $H_0=\gamma_1 D_1 \oplus (-\gamma_2 D_2) \oplus \gamma_3 D_3$
and $H_1$ has no zero mode. 
Along the path $t\in [0,1]$ the anticommutation relation
$\{H_t, \gamma_1\oplus(-\gamma_2)\oplus\gamma_3\}=0$ is kept.

The equivalence classes $[\mathcal{H},\gamma D,\gamma]$ form
an Abelian group $K^0(\{0\})$ as follows.
The sum is given by the direct sum
\begin{align}
 [\mathcal{H}_1,\gamma_1 D_1,\gamma_1]+[\mathcal{H}_2,\gamma_2 D_2,\gamma_2] = 
[\mathcal{H}_1\oplus \mathcal{H}_2,\gamma_1 D_1 \oplus \gamma_2 D_2,\gamma_1\oplus \gamma_2].
\end{align}
The identity element is given by $[\mathcal{H}_0,\gamma_0 D_0,\gamma_0]$
in which any of $\gamma_0 D_0$ has no zero mode.
The inverse element is given by $-[\mathcal{H},\gamma D,\gamma]=[\mathcal{H},-\gamma D,-\gamma]$.

It is known that $K^0(\{0\})\cong \mathbb{Z}$,
which is characterized by the Atiyah-Singer index of $D$.

By the suspension isomorphism, we can show that
\begin{equation}
 K^0(\{0\}) \cong K^1(I,\partial{I}).
\end{equation}
where $I$ is an interval in the parameter space 
and $\partial{I}$ is its two end points.
Parametrizing $s\in I=[-1,1]$, the isomorphic map is given by
\begin{equation}
 [\mathcal{H},\gamma D,\gamma] \mapsto [p^*\mathcal{H},\gamma(D+M_s)], 
\end{equation}
where the mass term $M_s$ satisfies $M_{-1}=M$, $M_{+1}=-M$ with a fixed value $M$
and $p^*\mathcal{H}$ is the pullback of the projection $I\to \{0\}$. 
Since it is just a copy of $\mathcal{H}$ at any value of $s$,
we simply ignore $p^*$ in the rest of this appendix.
Moreover, when $\mathcal{H}$ is obvious as in the lattice Hilbert space in this work, 
we neglect it and simply denote the group element only by the operator $\gamma(D+M_s)$.

Note that the definition of the $K^1(I,\partial{I})$
requires no chirality operator.
Besides, dimension of $\mathcal{H}$ can be either finite or infinite.
The massive lattice Wilson Dirac operator $\gamma (D_W+M_s)$ is then
naturally treated as the element of the same $K^1(I,\partial{I})$
group and we can compare the index with the one in continuum theory.

Defining the number of zero-crossing modes from positive to negative by $n_+$
  and that from negative to positive by $n_-$, the spectral flow
or equivalently the $\eta$ invariant,
\begin{equation}
n_+-n_-= - \frac{1}{2}\eta(\gamma(D_W-M))+\frac{1}{2}\eta(\gamma(D_W+M)),
\end{equation}
equals to the continuum Dirac operator index at sufficiently small lattice spacings.
For the standard choice of the Wilson term coefficient to be unity,
the second term can be neglected since it is trivially zero.

\section{$KO^0(I,\partial{I})$ and mod-two spectral flow}
\label{app:KO}

When the massless Dirac operator $D$ is real, 
it represents the $KO^{-1}(\{0\})$ group element.
For any eigenfunction $\phi$ satisfying $D\phi=i\lambda \phi$ with the eigenvalue $i\lambda$, 
$\phi^*$ is another eigenfunction with the oppositely signed eigenvalue $-i\lambda$ except for $\lambda=0$.
Therefore, the number of zero modes of $D$ mod 2  is a topological invariant called
the mod-two index that characterizes $KO^{-1}(\{0\})\cong \mathbb{Z}_2$.
The superscript ``$-1$'' of $KO^{-1}(\{0\})$ 
reflects removal of the chiral symmetry.
Thus, this index can be nontrivial in odd dimensions, 
such as the one in $SU(2)$ gauge theory in five dimensions.

Like the $K$ group case, the suspension isomorphism exists:
\begin{equation}
 KO^{-1}(\{0\}) \cong KO^0(I,\partial{I}),
\end{equation}
with the isomorphic map 
\begin{equation}
 [\mathcal{H},D] \mapsto [\mathcal{H}\oplus \mathcal{H}, \tau_1 \otimes D -i\tau_2 \otimes M_s,\tau_3\otimes 1], 
\end{equation}
where $\tau_{1,2,3}$ are the  Pauli matrices (in the flavor space)
and $M_s$ is a real mass parameter family taking $s\in [-1,1]$
which satisfies $M_{-1}=M$ and $M_{+1}=-M$.

There is another type of the mod-two index  when
an additional symmetry makes the zero modes to always appear in pairs.
The number of such zero pairs is a topological invariant.
In this case the Dirac operator represents an element of $KO^{-2}(\{0\})\cong \mathbb{Z}_2$
and there is also a suspension isomorphism
\begin{equation}
 KO^{-2}(\{0\})\cong KO^{-1}(I,\partial{I}).
\end{equation}
In fact, the above element $[\mathcal{H}\oplus \mathcal{H}, \tau_1 \otimes D -i\tau_2 \otimes M_s,\tau_3\otimes 1]$ 
can be understood as the one of $KO^{-1}(I,\partial{I})$
since we can forget the chirality operator $\tau_3\otimes 1$.

For the obtained massive Dirac operator $\tau_1 \otimes D -i\tau_2 \otimes M_s$,
every zero crossing occurs in a pair of
one from negative to positive and another from positive to negative.
Therefore, the standard index is trivially zero.
But the number of such pair-zero-crossings $n$ mod 2 is a topological invariant.
When the massive Wilson Dirac operator is real,
it can be identified as the $KO^0(I,\partial{I})$ or $KO^{-1}(I,\partial{I})$ group element,
and the mod-two spectral flow 
\begin{equation}
 n = \frac{1-\mbox{sgn}\det[(D_W+M_{+1})/(D_W+M_{-1})]}{2}\;\;\;\mbox{mod 2},
\end{equation}
equals to the mod-two index of the
original massless Dirac operator in the continuum limit.
When the coefficient of the Wilson term is unity, 
the determinant of $D_W+M_{-1}$ is always 1 and can be neglected.

\section{The APS index and gluing property}
\label{app:APS}

In this work, we consider the cases where
$M_s$ changes its sign by $-sM$ only in a limited region $T_-$,
while it is fixed at $M_s=+M$ in the rest of 
the region $T_+=T^{2n}-T_-$.
At $s=1$, the mass term has a domain-wall between
$T_+$ and $T_-$.
The Wilson Dirac operator with this domain-wall mass
term can still be interpreted as an element of $K^1(I,\partial I)$
or $KO^0(I,\partial{I})$ under some reasonable conditions including
that the Dirac operator is gapped at $s=\pm 1$.
But there is no simple isomorphism to the $K^0$ or $KO^{-1}$ groups.

In \cite{Fukaya:2019qlf,Fukaya:2020tjk}
it was proved that the corresponding spectral flows
equal to the APS index of the massless Dirac operator on $T_-$
with a non-local boundary condition imposed on $\partial T_-$,
and its mod-two version, respectively.

In the original definition, the APS index
has the so-called gluing property. 
Suppose we have an APS index $I_+$ on a manifold $X_+$
with a boundary $Y$ and another APS index $I_-$
on a manifold $X_-$ sharing the same boundary $Y$
but with the opposite orientation.
Then $I_++I_-$ equals to the AS index on
the glued closed manifold $X=X_+ \cup X_-$.
In our formulation, this property is viewed as follows.
Let us extend the path 
$s\in [1,2]$ where the mass in the $T_+$ region
vary as $M_s=(-2s+3)M$ while it is fixed at $M_s=-M$
in the $T_-$ region.
The second spectral flow along the extended path
equals to the APS index in the $T_+$ region.
From  the homotopic property of the joint
spectral flow $s\in [0,2]$, the sum of the
two APS indices on $T_-$ and $T_+$
must be equal to the AS index on the whole $T^{2n}$.
Thus, the gluing property of the APS index
is naturally contained in our $K$-theoretic formulas.

\end{appendix}